\pdfoutput=1
\documentclass[aps,arxiv,twocolumn,showpacs,superscriptaddress]{revtex4-1}  
\usepackage{graphicx}
\usepackage{amsmath}
\usepackage{amssymb}
\usepackage{mathtools}
\usepackage[version=4]{mhchem}

\usepackage{microtype}
\usepackage{hyperref}
\hypersetup{
  pdfauthor={Nongnuch Artrith},
  pdftitle={Efficient and Accurate Machine-Learning Interpolation of
    Atomic Energies in Compositions with Many Species}}

\begin{document}

\title{Efficient and Accurate Machine-Learning Interpolation of Atomic
  Energies in Compositions with Many Species}

\author{Nongnuch Artrith}
\email{nartrith@berkeley.edu}
\affiliation{%
  Department of Materials Science and Engineering,
  University of California, Berkeley, CA, USA}
\author{Alexander Urban}
\affiliation{%
  Department of Materials Science and Engineering,
  University of California, Berkeley, CA, USA}
\author{Gerbrand Ceder}
\email{gceder@berkeley.edu}
\affiliation{%
  Department of Materials Science and Engineering,
  University of California, Berkeley, CA, USA}
\affiliation{%
  Materials Science Division, Lawrence Berkeley National
  Laboratory, Berkeley, CA, USA}
\date{\today}

\begin{abstract}
  Machine-learning potentials (MLPs) for atomistic simulations are a
  promising alternative to conventional classical potentials.
  Current approaches rely on descriptors of the local atomic environment
  with dimensions that increase quadratically with the number of
  chemical species.
  In this article, we demonstrate that such a scaling can be avoided in
  practice.
  We show that a mathematically simple and computationally efficient
  descriptor with constant complexity is sufficient to represent
  transition-metal oxide compositions and biomolecules containing
  11~chemical species with a precision of around 3~meV/atom.
  This insight removes a perceived bound on the utility of MLPs and
  paves the way to investigate the physics of previously inaccessible
  materials with more than ten chemical species.
\end{abstract}

\maketitle


Atomic interaction potentials based on the interpolation of
\emph{first-principles} calculations with machine-learning algorithms
have the potential to enable efficient linear-scaling atomistic
simulations with an accuracy that is close to the reference
method~\cite{cpl395-2004-210, prl98-2007-146401, prl104-2010-136403,
  prl108-2012-058301}.
Such machine-learning potentials (MLPs) establish a relationship between
a unique descriptor and the total or atomic energy using, e.g.,
artificial neural networks (ANNs)~\cite{MontavonOrrMueller2012} or
Gaussian process regression (GPR, Kriging)~\cite{Rasmussen2006}.
However, the combined space of atomic coordinates and chemical species
grows rapidly with the number of chemical species, resulting in a formal
corresponding growth of the descriptor complexity and thus the
complexity of the MLP.
This scaling has so far limited current MLP approaches to compositions
with only a few chemical species~\cite{prb83-2011-153101,
  nl14-2014-2670, cms110-2015-20, pnas113-2016-8368} or atomic
structures~\cite{prl117-2016-135502}.
Overcoming this limitation is a very active field of
research~\cite{arxiv-1704.06439, pccp18-2016-13754}.

\begin{figure*}[t]
  \centering
  \includegraphics[width=0.95\textwidth]{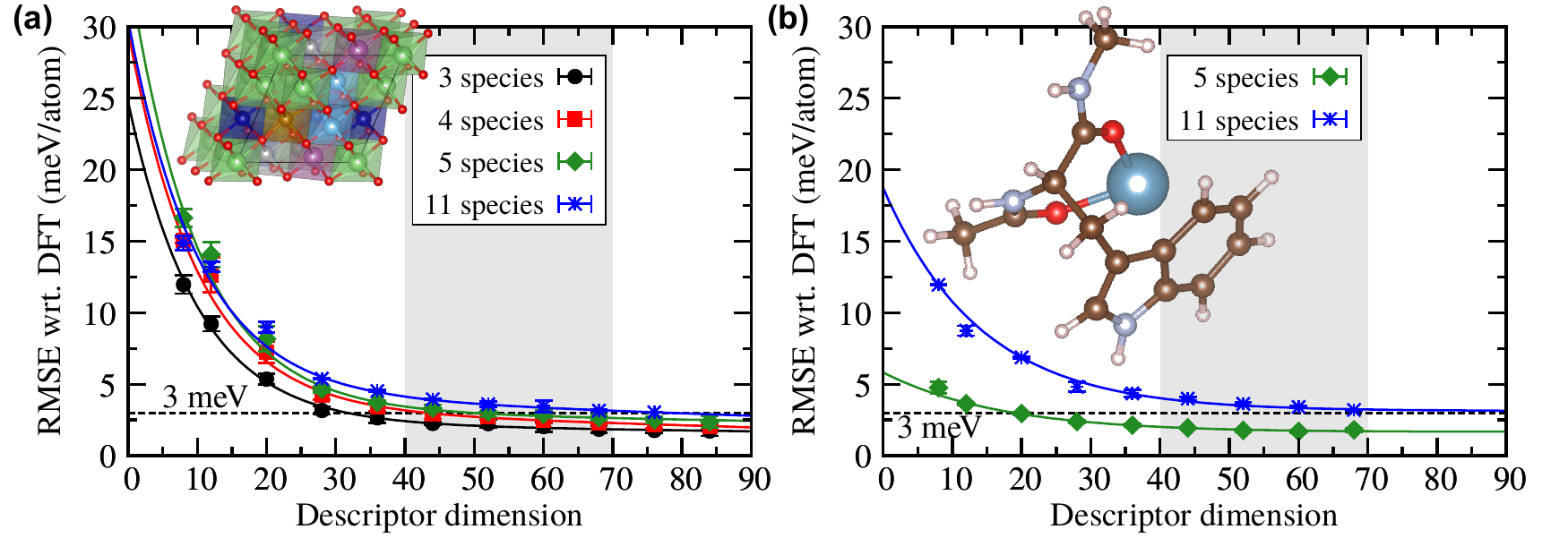}
  \caption{\label{fig:complexity}%
    Precision of artificial neural network (ANN) potentials as function
    of the dimension of the descriptor used to represent the local
    atomic environment.
    \textbf{(a)}~Root mean squared error (RMSE) of the ANN potential
    energies relative to their DFT references for \ce{Li$M$O2} systems with
    increasing number of chemical species: 3~species ($M$\,=\,Ti; black
    circles), 4~species ($M$\,=\,Ti, Ni; red squares), 5~species
    ($M$\,=\,Ti, Mn, Ni; green diamonds), and 11~species ($M$\,=\,Sc, Ti, V,
    Cr, Mn, Fe, Co, Ni, Cu; blue stars).
    The unit cell of a representative \ce{Li$M$O2}
    structure from the data set is shown as inset.
    \textbf{(b)}~An equivalent analysis for a data set with
    conformations
    of the 20~proteinogenic amino acids (5~chemical
    species: H, C, N, O, S; green diamonds) and their complexes with the
    divalent cations \ce{Ba^2+}, \ce{Ca^2+}, \ce{Cd^2+}, \ce{Hg^2+},
    \ce{Pb^2+}, \ce{Sr^2+} (in total 11~species; blue stars).
    The inset shows one conformation of a tryptophan dipeptide complex
    with \ce{Ca^2+}.
    Generally, the RMSE was evaluated after 3,000~training
    iterations, except for the two 11-species systems for which
    5,000~iterations were required.
    The error bars indicate the standard deviation of three
    independently trained ANN potentials, the gray region highlights
    descriptors that result in essentially converged ANN potentials with
    RMSE values around 3~meV/atom, and the lines are meant to guide the
    eye.}

\end{figure*}
In this article we demonstrate that the computational complexity of MLPs
does not necessarily grow with the number of chemical species, so that
MLPs for materials with ten or more chemical species are in principle
feasible and computationally efficient.
We show that, contrary to intuition and common belief, the same model
complexity that is optimal for a ternary material is also sufficient to
describe a system with 11 chemical species
(\textbf{Fig.~\ref{fig:complexity}}).
To illustrate these concepts, we consider two different materials
classes of practical relevance: cation-disordered lithium
transition-metal (TM) oxides, which have recently attracted interest as
high-energy-density cathode materials for Li-ion
batteries~\cite{sci343-2014-519, pnas112-2015-7650}, and proteinogenic
amino acids, i.e., the building blocks of proteins and their complexes
with divalent cations~\cite{sd3-2016-160009, sr6-2016-35772}.
We show that both of these high-dimensional materials systems can be
accurately modeled using MLPs based on a mathematically simple and
computationally efficient descriptor with constant complexity that we
will introduce in the following.

In the present work, we focus on MLPs that express the total structural
energy as the sum of atomic energy contributions and are in this respect
similar to other many-body potentials such as embedded atom
models~\cite{prB29-1984-6443, msr9-1993-251}.
However, unlike conventional potentials, the atomic energy is not
confined to a rigid functional form, but is represented by a flexible
non-linear machine-learning model that is trained to a descriptor of the
\emph{local atomic environment}.
In this context, the local atomic environment
$\sigma_{i}^{R_{\textup{c}}} \subset \sigma$ of an atom $i$ in a
structure $\sigma$ is defined as the \emph{local structure} given by the
set of coordinates $\{\mathbf{R}_{1}, \mathbf{R}_{2}, \ldots\}$ of all
atoms within a cutoff distance $R_{\textup{c}}$ from atom $i$ and the
\emph{local composition}, i.e., the corresponding chemical species
$\{t_{1}, t_{2}, \ldots\}$.
To be physically meaningful and transferable between equivalent
structures, the descriptor needs to be invariant with respect to
translation and rotation of the structure and the exchange of equivalent
atoms.
Several transformations for $\sigma_{i}^{R_{\textup{c}}}$ into invariant
representations $\widetilde{\sigma}_{i}^{R_{\textup{c}}}$ have been
proposed in the literature~\cite{jcp134-2011-074106, prB87-2013-184115,
  jcp139-2013-184118, prB89-2014-205118, ijqc115-2015-1032,
  ijqc115-2015-1094, ijqc115-2015-1084}, and the most commonly used
methods for MLPs are the \emph{symmetry functions} by Behler and
Parrinello (BP)~\cite{prl98-2007-146401, jcp134-2011-074106} and the
\emph{smooth overlap of atomic positions} (SOAP) approach by Bart\'ok,
Kondor, and Cs\'anyi~\cite{prB87-2013-184115, ijqc115-2015-1051,
  pccp18-2016-13754}.
With an invariant descriptor $\widetilde{\sigma}_{i}^{R_{\textup{c}}}$,
the total MLP energy of a structure $\sigma$ can then be expressed as
\[
  E(\sigma)
  = \sum_{i}^{\textup{atoms}}
    \textup{MLP}_{t_{i}}(\widetilde{\sigma}_{i}^{R_{\textup{c}}})
\].

Our approach draws inspiration from the strength of the established
descriptor methods but explicitly maintains the distinction between
local \emph{structure} and \emph{composition} by using two sets of
invariant coordinates,
$^{\{\mathbf{R}\}}\widetilde{\sigma}_{i}^{R_{\textup{c}}}$ and
$^{\{t\}}\widetilde{\sigma}^{R_{\textup{c}}}_{i}$, that separately
encode the atomic positions and species.
The union of both sets,
$\widetilde{\sigma}_{i}^{R_{\textup{c}}} =
{^{\{\mathbf{R}\}}\widetilde{\sigma}}_{i}^{R_{\textup{c}}} \cup
{^{\{t\}}\widetilde{\sigma}}^{R_{\textup{c}}}_{i}$, is used as a
combined descriptor for an ANN-based MLP (ANN potential).
As \emph{structural} descriptor
${^{\{\mathbf{R}\}}\widetilde{\sigma}}_{i}^{R_{\textup{c}}}$ we choose the
expansion coefficients of the radial (bond length) and angular (bond
angle) distribution functions in a complete basis set
$\{\phi_{\alpha}\}$,
\begin{align}
  \mathrm{RDF}_{i}(r)
  &= \sum_{\alpha} c^{(2)}_{\alpha}
    \phi_{\alpha}(r)
    \quad\text{for}\quad{}0\leq{}r\leq{}R_{\textup{c}}
  \label{eq:RDF-expansion}
\\
  \mathrm{ADF}_{i}(\theta)
  &= \sum_{\alpha} c^{(3)}_{\alpha}
    \phi_{\alpha}(\theta)
    \quad\text{for}\quad{}0\leq{}\theta\leq{}\pi
  \label{eq:ADF-expansion}
  \quad ,
\end{align}
and the \emph{compositional} descriptor
${^{\{t\}}\widetilde{\sigma}}^{R_{\textup{c}}}_{i}$ is given by the
expansion coefficients of the same distribution functions but with
atomic contributions that are weighted differently for each chemical
species.
The RDF and ADF obey the invariants of the atomic energy, and basing the
descriptor on an expansion in a complete basis set allows its systematic
refinement by converging the number of basis functions.
We implemented the descriptor into the free and open-source \emph{atomic
  energy network} package~\cite{cms114-2016-135}.

In general, multi-layer ANNs can reproduce any function with arbitrary
precision~\cite{nn4-1991-251}.
However, the resolution of the invariant descriptor determines the
maximal precision with which an ANN potential can resolve the chemical
space of a given material.
To determine the resolution of our combined descriptor, we trained ANN
potentials to extensive reference data sets with different numbers of
chemical species.
We consider the resolution satisfactory if the ANN potential can
reproduce the reference energies of our data sets with a precision of
$\sim$3~meV/atom, which is the order of magnitude of the noise in our
reference data.

\textbf{Figure~\ref{fig:complexity}a} shows the precision that can be
achieved in representing Li-TM oxides with different numbers of TM
species using ANN potentials based on the combined descriptor with
different numbers of basis functions.
The reference set for the ANN potential training comprised Hubbard-U
corrected~\cite{prB52-1995-R5467, jpcm9-1997-767, cms50-2011-2295}
density-functional theory (DFT) energies and optimized structures of
16,047~\ce{Li$M$O2} configurations in the rocksalt structure with
different compositions based on nine TMs (Sc, Ti, V, Cr, Mn, Fe, Co, Ni,
and Cu) and cation arrangements with up to 36~atoms.
For all DFT+U calculations we employed the PBE exchange-correlation
functional~\cite{prl77-96-3865} with projector-augmented
wave~\cite{prB50-1994-17953} pseudopotentials as implemented in
VASP~\cite{prB54-1996-11169, cms6-1996-15}.
DFT energies and atomic forces were converged to 0.05~meV per atom and
50~meV/\AA{}, respectively, gamma-centered k-point meshes with a density
of 1000~divided by the number of atoms were used, and the plane-wave
cutoff was 520~eV.
VASP input files were generated using the pymatgen software with default
parameters~\cite{cms68-2013-314}.
Structures with up to 5~chemical species were generated by systematic
enumeration, and random atomic configurations were generated for
compositions with 6--11~chemical species.
Further information about the generation of these reference structures,
the parameters of our DFT calculations, and the architecture of the ANNs
are given in the Appendix.

As seen in \textbf{Fig.~\ref{fig:complexity}a}, the ANN potentials
achieve a root mean squared error (RMSE) of $\sim$3~meV/atom relative to
the DFT reference energies with a descriptor dimension of 44 (i.e.,
22~basis functions).
Note that, for the present work, we employed the same number of basis
functions for the radial and angular expansion (i.e., 11~each), though
this is not a general requirement of the methodology.
Increasing the descriptor dimension beyond 52~or~60 results in a minor
additional reduction of the RMSE at the cost of significantly increased
computational effort.
We emphasize that this RMSE is purely a quality measure of the
descriptor precision and does not reflect the accuracy of the ANN
potentials in simulations, which would have to be carefully validated
separately.

The RMSE was evaluated after 3,000~training iterations using the LM-BFGS
method~\cite{siam16-1995-1190, toms23-1997-550}, however, with
increasing number of species and increasing descriptor size the required
number of training iterations to achieve convergence generally also
increases.
Thus, the ANN potentials for 11~chemical species and descriptor
dimensions above 40 have not converged after 3,000~iterations, and the
RMSEs after 5,000~iterations are shown in
\textbf{Fig.~\ref{fig:complexity}}.
The unconverged RMSE after 3,000~training iterations is shown in
\textbf{Fig.~S2} in the Appendix.

Remarkably, the optimal descriptor dimension is essentially independent
of the number of chemical species in the composition, and a descriptor
dimension of 44 is sufficient to capture the structural and chemical
features of the distinct atomic configurations in the \ce{Li$M$O2} data
set with up to 11~chemical species.

\textbf{Figure~\ref{fig:complexity}b} shows the equivalent analysis for
the first-principles energies and structures of 45,892 conformations of
the proteinogenic amino acids (5~chemical species: H, C, N, O, and S)
and their complexes with the six divalent cations \ce{Ba^2+},
\ce{Ca^2+}, \ce{Cd^2+}, \ce{Hg^2+}, \ce{Pb^2+}, \ce{Sr^2+} (a total of
11~chemical species) by Ropo, Schneider, Baldauf, and
Blum~\cite{sd3-2016-160009} based on DFT calculations
(PBE+TS-vdW~\cite{prl102-09-073005}) using the \textsc{FHI-aims}
package~\cite{cpc180-2009-2175}.
This data set was compiled specifically for the parametrization of
atomic potentials and thoroughly samples the relevant conformational
space~\cite{sd3-2016-160009}, an important first step towards improved
force fields for proteins~\cite{coisb24-2014-98}.
The high precision of the ANN potentials with an RMSE of
$\sim$3~meV/atom for 5~and~11 chemical species indicates that our
combined descriptor is not limited to crystal structures with similar
atomic positions, but is also suitable to distinguish between continuous
atomic arrangements.

\begin{figure}[tb]
  \centering
  \includegraphics[width=0.95\columnwidth]{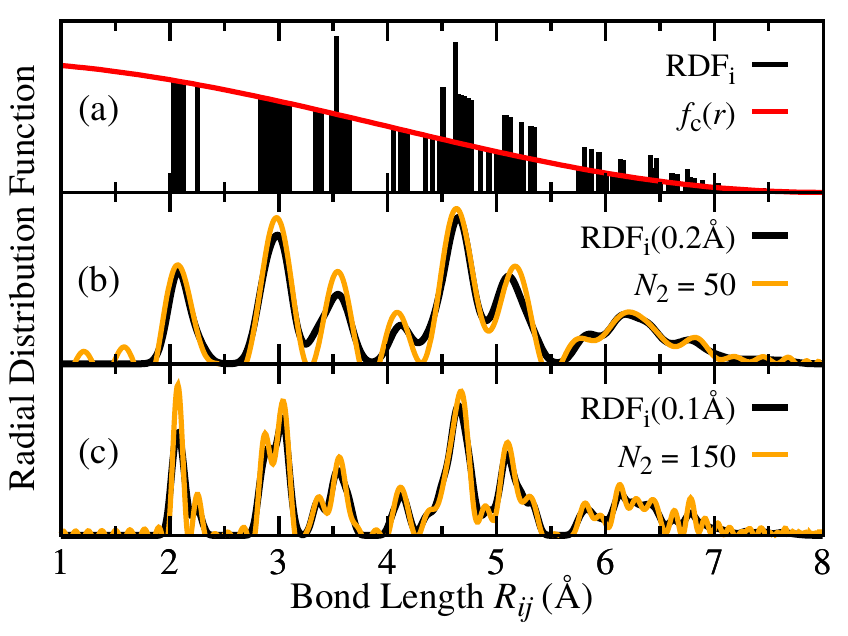}
  \caption{\label{fig:RDF-convergence}%
    \textbf{(a)}~Discrete atom-centered radial distribution function
    ($\textup{RDF}_{i}$) for a lithium site in a structure with
    composition Li$_{2}$MnNiO$_{4}$ (black lines) and the cosine cutoff
    function $f_{\textup{c}}$ for a cutoff radius of
    $R_{\textup{c}}=8$\,\AA{}. \textbf{(b)}~Convolution of the RDF
    of~(a) with a Gaussian function with a width of 0.2\,\AA{} (black
    line) and the reconstructed RDF from a Chebyshev expansion with a
    radial order $N_{2}=50$ (orange line).  \textbf{(c)}~Same as~(b),
    but with a Gaussian width of 0.1\,\AA{} and an expansion order of
    $N_{2}=150$.}
\end{figure}
To understand the significance of these observations, we first describe
the details of the structural and compositional descriptor.
We begin by expressing the atom-centered radial and angular distribution
functions of Eqs.~\eqref{eq:RDF-expansion} and~\eqref{eq:ADF-expansion}
in terms of discrete delta functions centered at the bond lengths
between atoms $j$ and the central atom $i$,
$R_{ij}=||\mathbf{R}_{j}-\mathbf{R}_{i}||$, and the bond angle
$\theta_{ijk}=\angle(\mathbf{R}_{j}-\mathbf{R}_{i},
\mathbf{R}_{k}-\mathbf{R}_{i})$
\begin{align}
  \mathrm{RDF}_{i}(r)
  &= \sum_{\mathclap{\;\;\mathbf{R}_{j}\in\,\sigma_{i}^{R_{\textup{c}}}}}
    \delta(r - R_{ij}) \,f_{\textup{c}}(R_{ij}) \,w_{t_{j}}
  \label{eq:RDF}
\\
  \mathrm{ADF}_{i}(\theta)
  &= \sum_{\mathclap{\;\;\mathbf{R}_{j},\mathbf{R}_{k}\in\,\sigma_{i}^{R_{\textup{c}}}}}
    \delta(\theta - \theta_{ijk})\,f_{\textup{c}}(R_{ij})\,f_{\textup{c}}(R_{ik})
    \,w_{t_{j}}w_{t_{k}}
  \quad ,
  \label{eq:ADF}
\end{align}
where $f_{\textup{c}}$ is a cutoff function that smoothly goes to zero
at $R_{\textup{c}}$ (in practice, we use
$f_{\textup{c}}(r) = 0.5[\cos(r\cdot\pi/R_{\textup{c}}) + 1]$).
The weights $w_{t_{j}}$ and $w_{t_{k}}$ are~1 for the \emph{structural
  descriptor} $^{\{\mathbf{R}\}}\widetilde{\sigma}_{i}^{R_{\textup{c}}}$
and take on species-dependent values for the \emph{compositional
  descriptor} $^{\{t\}}\widetilde{\sigma}_{i}^{R_{\textup{c}}}$.
Here, we followed the (Ising-model) pseudo-spin convention commonly used
for lattice models~\cite{pa128-1984-334}, i.e.,
$w_{l} = 0, \pm{}1, \pm{}2, \ldots$ where 0 is omitted for even numbers of
species.
For the expansions Eqs.~\eqref{eq:RDF-expansion}
and~\eqref{eq:ADF-expansion} we choose a complete orthonormal basis
$\{\phi_{\alpha}\}$, i.e., $\int\phi_{\alpha}\overline{\phi}_{\beta}=1$
if $\alpha=\beta$ and $0$ else.
With this choice, the expansion coefficients are given by
\begin{align}
  c^{(2)}_{\alpha}
  &= \sum_{\mathclap{\;\;\mathbf{R}_{j}\in\,\sigma_{i}^{R_{\textup{c}}}}}
     \phi_{\alpha}(R_{ij}) \,f_{\textup{c}}(R_{ij}) \,w_{t_{j}}
    \qquad\text{and}
  \label{eq:coeff-(2)}
\\
  c^{(3)}_{\alpha}
  &= \sum_{\mathclap{\;\;\mathbf{R}_{j},\mathbf{R}_{k}\in\,\sigma_{i}^{R_{\textup{c}}}}}
     \phi_{\alpha}(\theta_{ijk}) \,f_{\textup{c}}(R_{ij})\,f_{\textup{c}}(R_{ik})
    \,w_{t_{j}}w_{t_{k}}
  \label{eq:coeff-(3)}
  \quad .
\end{align}
A derivation of Eqs.~\eqref{eq:coeff-(2)} and~\eqref{eq:coeff-(3)} can
be found in the Appendix.
The expansions are truncated at finite radial and angular orders $N_{2}$
and $N_{3}$ that determine the dimension (i.e., the complexity) and the
resolution of the descriptor, i.e.,
$^{\{\mathbf{R}\}}\widetilde{\sigma}_{i}^{R_{\textup{c}}}=\{{^{\{\mathbf{R}\}}c^{(2)}_{1}},
\ldots, {^{\{\mathbf{R}\}}c^{(2)}_{N_{2}}},
{^{\{\mathbf{R}\}}c^{(3)}_{1}}, \ldots,
{^{\{\mathbf{R}\}}c^{(3)}_{N_{3}}}\}$.

For this article, we employed the Chebyshev polynomials of the first
kind as basis functions (see Appendix), as they can be defined in terms of a
recurrence relation that allows for highly efficient numerical
evaluation of the function values and their derivatives.
With this choice of basis functions,
\textbf{Fig.~\ref{fig:RDF-convergence}} shows the RDF as reconstructed
based on the structural expansion coefficients
$\{^{\{\mathbf{R}\}}c_{\alpha}^{(2)}\}$ for two different orders
($N_{2}=50$ and $N_{2}=150$).
From comparison with Gaussian convolutions of the discrete RDF, the
radial resolution of the expansion order $N_{2}=150$ is
around~0.1\,\AA{}.
Atomic features on smaller scales may affect the shape of the RDF but do
not give rise to distinct peaks.
The expansion of the ADF is completely analogous.

%
We note that the radial and angular BP symmetry
functions~\cite{prl98-2007-146401, jcp134-2011-074106} can be cast into
the form of Eqs.~\eqref{eq:coeff-(2)} and~\eqref{eq:coeff-(3)} but are
neither orthogonal nor systematically refinable.
The relationship of our structural descriptor to the coefficients of a
basis set expansion is, in turn, closer in spirit to the SOAP
method~\cite{prl104-2010-136403, prB87-2013-184115} which is based on
the power spectrum of the atomic density of the local atomic
environment.
SOAP allows for a rigorous and systematic description of the local
structure, which comes at the cost of an arithmetically (and
computationally) more complex formalism.
However, by limiting the descriptor to radial and angular contributions
our method maintains the simple analytic nature of the BP approach that
allows for a highly efficient numerical implementation and
straightforward differentiation (which is required for the calculation
of analytic forces and higher derivatives).
Basing the radial and angular descriptors on an expansion in a complete
basis set allows their systematic refinement in the spirit of the SOAP
approach, though our approach is limited to two- and three-body
interactions.

Also note that decomposing the local atomic environment into $n$-body
contributions as done in our structural descriptor is an established and
well-tested approach for lattice models such as the cluster expansion
(CE) method~\cite{ssp47-1994-33, cms1-1993-144}.
In CE models, the total configurational energy is expanded in a basis
set consisting of site clusters $(s_{i}s_{j}\ldots)$ with increasing
numbers of lattice sites $s_{\alpha}$, i.e., point clusters, pairs,
trimers, \ldots, $n$-tuples.
The site clusters form a complete basis set, and the configurational
averages of all equivalent clusters (the cluster correlations) are the
descriptor of the CE model.
Unlike MLPs, the CE energy is a \emph{linear} function of the
descriptor.
For the case of the continuous structural energy,
Thompson~\textit{et~al.}  demonstrated that a linear potential based on
SOAP (which also is a complete basis of the local structure) can achieve
reasonable accuracy in practice if a sufficient number of basis
functions is used~\cite{jocp-2014-SNAP}.

However, the strength of non-linear machine-learning models is that they
do not require mathematically complete descriptors as long as the
descriptor is able to differentiate between all \emph{relevant} samples.
This property is exploited, for example, in the area of image
recognition and text classification~\cite{RogatiYang2002}.
In practice this means that even an incomplete descriptor of the local
atomic environment may be sufficient to construct a non-linear MLP if
that descriptor is able to differentiate between all \emph{relevant}
local atomic structures, i.e., the descriptor does not have to resolve
all hypothetically possible sets of three dimensional coordinates.

\begin{figure}[tb]
  \centering
  \includegraphics[width=0.95\columnwidth]{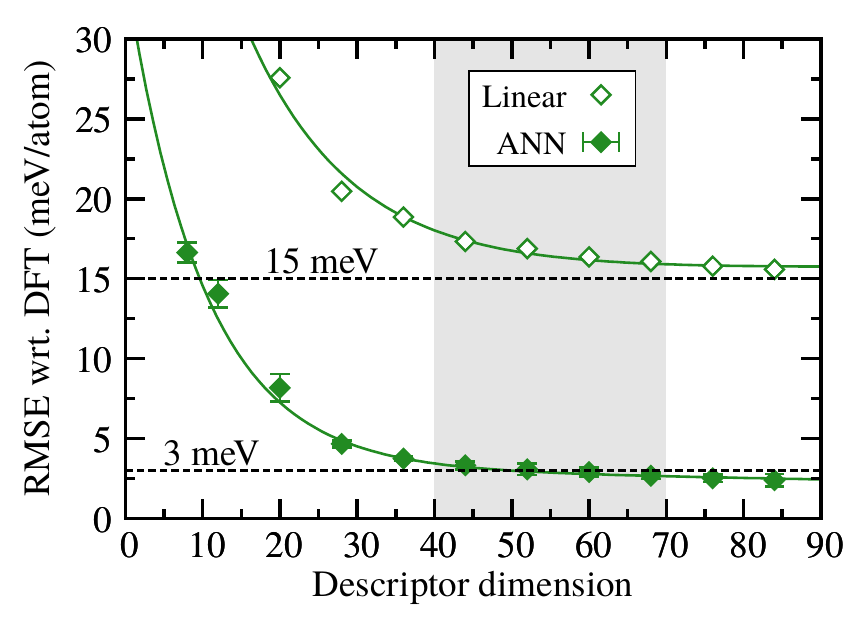}
  \caption{\label{fig:linear-vs-ANN}%
    Convergence of the root mean squared error (RMSE) of the predicted
    \ce{Li$M$O2} (5~species, $M$\,=\,Ti, Mn, Ni) energy for a linear
    model (empty diamonds) and a non-linear ANN (filled diamonds) with
    the dimension of the combined descriptor.  The gray region
    highlights descriptors that result in essentially converged ANN
    potentials with RMSE values around 3~meV/atom.  The lines are meant
    to guide the eye.}
\end{figure}
This behavior is exemplified in \textbf{Fig.~\ref{fig:linear-vs-ANN}},
which compares the precision of an ANN potential for the \ce{Li$M$O2}
data set with 5~chemical species (10,175~atomic configurations) with
that of a linear energy model as function of the descriptor dimension.
As seen in the figure, the ANN achieves an RMSE of $\leq$3~meV/atom with
descriptor dimensions of 44 (22~basis functions) and larger.
Comparison with \textbf{Fig.~\ref{fig:RDF-convergence}} shows that such
a small basis set corresponds to a coarse representation of the RDF and
the ADF, however, obviously this level of approximation is sufficient
for the ANN potential to differentiate between all structural and
compositional features in the reference set.
This is not the case for the linear model whose RMSE is $>$15~meV/atom
even for a descriptor dimension of 84.

In conclusion, we showed that machine-learning potentials do not require
(mathematically) complete descriptors of the local atomic environment to
reproduce potential energy surfaces with high precision.
With this insight, we devised a combined descriptor of the local atomic
\emph{structure} and \emph{composition} whose complexity does not scale
with the number of chemical species.
The method is conceptually simple and allows for highly efficient
numerical implementations.
The utility of the approach was demonstrated for two exemplary materials
classes, lithium transition-metal oxides and amino acid complexes, each
separately comprising compositions with 11 different chemical species.
We showed that the potential energy landscape of both example systems
can be represented with high precision by artificial neural network
potentials using the combined descriptor achieving a resolution of
around 3~meV/atom.
Hence, machine-learning potentials are in practice not limited to
compositions with small numbers of chemical species as previously argued
in the literature and may be effective for the modeling of
high-dimensional materials such as oxide solid solutions and peptide
chains.

\newpage
\section{Acknowledgments}

This work was supported by the Office of Naval Research (ONR) under
ONR award N00014-14-1-0444.
This work used mainly the computational facilities of the Extreme
Science and Engineering Discovery Environment (XSEDE), which is
supported by National Science Foundation grant no. ACI-1053575.
Additional computational resources from the University of California
Berkeley, HPC Cluster (SAVIO) are also gratefully acknowledged.


\bibliography{Artrith-Bibliography.bib}

\begin{thebibliography}{52}%
\makeatletter
\providecommand \@ifxundefined [1]{%
 \@ifx{#1\undefined}
}%
\providecommand \@ifnum [1]{%
 \ifnum #1\expandafter \@firstoftwo
 \else \expandafter \@secondoftwo
 \fi
}%
\providecommand \@ifx [1]{%
 \ifx #1\expandafter \@firstoftwo
 \else \expandafter \@secondoftwo
 \fi
}%
\providecommand \natexlab [1]{#1}%
\providecommand \enquote  [1]{``#1''}%
\providecommand \bibnamefont  [1]{#1}%
\providecommand \bibfnamefont [1]{#1}%
\providecommand \citenamefont [1]{#1}%
\providecommand \href@noop [0]{\@secondoftwo}%
\providecommand \href [0]{\begingroup \@sanitize@url \@href}%
\providecommand \@href[1]{\@@startlink{#1}\@@href}%
\providecommand \@@href[1]{\endgroup#1\@@endlink}%
\providecommand \@sanitize@url [0]{\catcode `\\12\catcode `\$12\catcode
  `\&12\catcode `\#12\catcode `\^12\catcode `\_12\catcode `\%12\relax}%
\providecommand \@@startlink[1]{}%
\providecommand \@@endlink[0]{}%
\providecommand \url  [0]{\begingroup\@sanitize@url \@url }%
\providecommand \@url [1]{\endgroup\@href {#1}{\urlprefix }}%
\providecommand \urlprefix  [0]{URL }%
\providecommand \Eprint [0]{\href }%
\providecommand \doibase [0]{http://dx.doi.org/}%
\providecommand \selectlanguage [0]{\@gobble}%
\providecommand \bibinfo  [0]{\@secondoftwo}%
\providecommand \bibfield  [0]{\@secondoftwo}%
\providecommand \translation [1]{[#1]}%
\providecommand \BibitemOpen [0]{}%
\providecommand \bibitemStop [0]{}%
\providecommand \bibitemNoStop [0]{.\EOS\space}%
\providecommand \EOS [0]{\spacefactor3000\relax}%
\providecommand \BibitemShut  [1]{\csname bibitem#1\endcsname}%
\let\auto@bib@innerbib\@empty
\bibitem [{\citenamefont {Lorenz}\ \emph {et~al.}(2004)\citenamefont {Lorenz},
  \citenamefont {{Gro{\ss}}},\ and\ \citenamefont
  {{Scheffler}}}]{cpl395-2004-210}%
  \BibitemOpen
  \bibfield  {author} {\bibinfo {author} {\bibfnamefont {S.}~\bibnamefont
  {Lorenz}}, \bibinfo {author} {\bibfnamefont {A.}~\bibnamefont {{Gro{\ss}}}},
  \ and\ \bibinfo {author} {\bibfnamefont {M.}~\bibnamefont {{Scheffler}}},\
  }\href {\doibase 10.1016/j.cplett.2004.07.076} {\bibfield  {journal}
  {\bibinfo  {journal} {Chem. Phys. Lett.}\ }\textbf {\bibinfo {volume}
  {395}},\ \bibinfo {pages} {210} (\bibinfo {year} {2004})}\BibitemShut
  {NoStop}%
\bibitem [{\citenamefont {Behler}\ and\ \citenamefont
  {Parrinello}(2007)}]{prl98-2007-146401}%
  \BibitemOpen
  \bibfield  {author} {\bibinfo {author} {\bibfnamefont {J.}~\bibnamefont
  {Behler}}\ and\ \bibinfo {author} {\bibfnamefont {M.}~\bibnamefont
  {Parrinello}},\ }\href {\doibase 10.1103/PhysRevLett.98.146401} {\bibfield
  {journal} {\bibinfo  {journal} {Phys. Rev. Lett.}\ }\textbf {\bibinfo
  {volume} {98}},\ \bibinfo {pages} {146401} (\bibinfo {year}
  {2007})}\BibitemShut {NoStop}%
\bibitem [{\citenamefont {Bart\'ok}\ \emph {et~al.}(2010)\citenamefont
  {Bart\'ok}, \citenamefont {Payne}, \citenamefont {Kondor},\ and\
  \citenamefont {Cs\'anyi}}]{prl104-2010-136403}%
  \BibitemOpen
  \bibfield  {author} {\bibinfo {author} {\bibfnamefont {A.~P.}\ \bibnamefont
  {Bart\'ok}}, \bibinfo {author} {\bibfnamefont {M.~C.}\ \bibnamefont {Payne}},
  \bibinfo {author} {\bibfnamefont {R.}~\bibnamefont {Kondor}}, \ and\ \bibinfo
  {author} {\bibfnamefont {G.}~\bibnamefont {Cs\'anyi}},\ }\href {\doibase
  10.1103/PhysRevLett.104.136403} {\bibfield  {journal} {\bibinfo  {journal}
  {Phys. Rev. Lett.}\ }\textbf {\bibinfo {volume} {104}},\ \bibinfo {pages}
  {136403} (\bibinfo {year} {2010})}\BibitemShut {NoStop}%
\bibitem [{\citenamefont {Rupp}\ \emph {et~al.}(2012)\citenamefont {Rupp},
  \citenamefont {Tkatchenko}, \citenamefont {M\"uller},\ and\ \citenamefont
  {von Lilienfeld}}]{prl108-2012-058301}%
  \BibitemOpen
  \bibfield  {author} {\bibinfo {author} {\bibfnamefont {M.}~\bibnamefont
  {Rupp}}, \bibinfo {author} {\bibfnamefont {A.}~\bibnamefont {Tkatchenko}},
  \bibinfo {author} {\bibfnamefont {K.-R.}\ \bibnamefont {M\"uller}}, \ and\
  \bibinfo {author} {\bibfnamefont {O.~A.}\ \bibnamefont {von Lilienfeld}},\
  }\href {\doibase 10.1103/PhysRevLett.108.058301} {\bibfield  {journal}
  {\bibinfo  {journal} {Phys. Rev. Lett.}\ }\textbf {\bibinfo {volume} {108}},\
  \bibinfo {pages} {058301} (\bibinfo {year} {2012})}\BibitemShut {NoStop}%
\bibitem [{\citenamefont {Montavon}\ \emph {et~al.}(2012)\citenamefont
  {Montavon}, \citenamefont {Orr},\ and\ \citenamefont
  {M\"uller}}]{MontavonOrrMueller2012}%
  \BibitemOpen
  \bibinfo {editor} {\bibfnamefont {G.}~\bibnamefont {Montavon}}, \bibinfo
  {editor} {\bibfnamefont {G.~B.}\ \bibnamefont {Orr}}, \ and\ \bibinfo
  {editor} {\bibfnamefont {K.-R.}\ \bibnamefont {M\"uller}},\ eds.,\ \href@noop
  {} {\emph {\bibinfo {title} {Neural Networks: Tricks of the Trade (Second
  Edition)}}},\ \bibinfo {series} {Lecture Notes in Computer Science}, Vol.\
  \bibinfo {volume} {7700}\ (\bibinfo  {publisher} {Springer Berlin
  Heidelberg},\ \bibinfo {year} {2012})\BibitemShut {NoStop}%
\bibitem [{\citenamefont {Rasmussen}\ and\ \citenamefont
  {Williams}(2006)}]{Rasmussen2006}%
  \BibitemOpen
  \bibfield  {author} {\bibinfo {author} {\bibfnamefont {C.~E.}\ \bibnamefont
  {Rasmussen}}\ and\ \bibinfo {author} {\bibfnamefont {C.~K.~I.}\ \bibnamefont
  {Williams}},\ }\href@noop {} {\emph {\bibinfo {title} {Gaussian Processes for
  Machine Learning}}}\ (\bibinfo  {publisher} {MIT University Press Group
  Ltd},\ \bibinfo {year} {2006})\BibitemShut {NoStop}%
\bibitem [{\citenamefont {Artrith}\ \emph {et~al.}(2011)\citenamefont
  {Artrith}, \citenamefont {Morawietz},\ and\ \citenamefont
  {Behler}}]{prb83-2011-153101}%
  \BibitemOpen
  \bibfield  {author} {\bibinfo {author} {\bibfnamefont {N.}~\bibnamefont
  {Artrith}}, \bibinfo {author} {\bibfnamefont {T.}~\bibnamefont {Morawietz}},
  \ and\ \bibinfo {author} {\bibfnamefont {J.}~\bibnamefont {Behler}},\ }\href
  {\doibase 10.1103/PhysRevB.83.153101} {\bibfield  {journal} {\bibinfo
  {journal} {Phys. Rev. B}\ }\textbf {\bibinfo {volume} {83}},\ \bibinfo
  {pages} {153101} (\bibinfo {year} {2011})}\BibitemShut {NoStop}%
\bibitem [{\citenamefont {Artrith}\ and\ \citenamefont
  {Kolpak}(2014)}]{nl14-2014-2670}%
  \BibitemOpen
  \bibfield  {author} {\bibinfo {author} {\bibfnamefont {N.}~\bibnamefont
  {Artrith}}\ and\ \bibinfo {author} {\bibfnamefont {A.~M.}\ \bibnamefont
  {Kolpak}},\ }\href {\doibase 10.1021/nl5005674} {\bibfield  {journal}
  {\bibinfo  {journal} {Nano Lett.}\ }\textbf {\bibinfo {volume} {14}},\
  \bibinfo {pages} {2670–} (\bibinfo {year} {2014})}\BibitemShut {NoStop}%
\bibitem [{\citenamefont {Artrith}\ and\ \citenamefont
  {Kolpak}(2015)}]{cms110-2015-20}%
  \BibitemOpen
  \bibfield  {author} {\bibinfo {author} {\bibfnamefont {N.}~\bibnamefont
  {Artrith}}\ and\ \bibinfo {author} {\bibfnamefont {A.~M.}\ \bibnamefont
  {Kolpak}},\ }\href {\doibase 10.1016/j.commatsci.2015.07.046} {\bibfield
  {journal} {\bibinfo  {journal} {Comput. Mater. Sci.}\ }\textbf {\bibinfo
  {volume} {110}},\ \bibinfo {pages} {20–28} (\bibinfo {year}
  {2015})}\BibitemShut {NoStop}%
\bibitem [{\citenamefont {Morawietz}\ \emph {et~al.}(2016)\citenamefont
  {Morawietz}, \citenamefont {Singraber}, \citenamefont {Dellago},\ and\
  \citenamefont {Behler}}]{pnas113-2016-8368}%
  \BibitemOpen
  \bibfield  {author} {\bibinfo {author} {\bibfnamefont {T.}~\bibnamefont
  {Morawietz}}, \bibinfo {author} {\bibfnamefont {A.}~\bibnamefont
  {Singraber}}, \bibinfo {author} {\bibfnamefont {C.}~\bibnamefont {Dellago}},
  \ and\ \bibinfo {author} {\bibfnamefont {J.}~\bibnamefont {Behler}},\ }\href
  {\doibase 10.1073/pnas.1602375113} {\bibfield  {journal} {\bibinfo  {journal}
  {Proc. Natl. Acad. Sci. USA}\ }\textbf {\bibinfo {volume} {113}},\ \bibinfo
  {pages} {8368} (\bibinfo {year} {2016})}\BibitemShut {NoStop}%
\bibitem [{\citenamefont {Faber}\ \emph {et~al.}(2016)\citenamefont {Faber},
  \citenamefont {Lindmaa}, \citenamefont {von Lilienfeld},\ and\ \citenamefont
  {Armiento}}]{prl117-2016-135502}%
  \BibitemOpen
  \bibfield  {author} {\bibinfo {author} {\bibfnamefont {F.~A.}\ \bibnamefont
  {Faber}}, \bibinfo {author} {\bibfnamefont {A.}~\bibnamefont {Lindmaa}},
  \bibinfo {author} {\bibfnamefont {O.~A.}\ \bibnamefont {von Lilienfeld}}, \
  and\ \bibinfo {author} {\bibfnamefont {R.}~\bibnamefont {Armiento}},\ }\href
  {\doibase 10.1103/PhysRevLett.117.135502} {\bibfield  {journal} {\bibinfo
  {journal} {Phys. Rev. Lett.}\ }\textbf {\bibinfo {volume} {117}},\ \bibinfo
  {pages} {135502} (\bibinfo {year} {2016})}\BibitemShut {NoStop}%
\bibitem [{\citenamefont {Huo}\ and\ \citenamefont
  {Rupp}(2017)}]{arxiv-1704.06439}%
  \BibitemOpen
  \bibfield  {author} {\bibinfo {author} {\bibfnamefont {H.}~\bibnamefont
  {Huo}}\ and\ \bibinfo {author} {\bibfnamefont {M.}~\bibnamefont {Rupp}},\
  }\href@noop {} {\bibfield  {journal} {\bibinfo  {journal} {arXiv}\ }
  (\bibinfo {year} {2017})},\ \Eprint {http://arxiv.org/abs/1704.06439}
  {1704.06439} \BibitemShut {NoStop}%
\bibitem [{\citenamefont {De}\ \emph {et~al.}(2016)\citenamefont {De},
  \citenamefont {Bartok}, \citenamefont {Csanyi},\ and\ \citenamefont
  {Ceriotti}}]{pccp18-2016-13754}%
  \BibitemOpen
  \bibfield  {author} {\bibinfo {author} {\bibfnamefont {S.}~\bibnamefont
  {De}}, \bibinfo {author} {\bibfnamefont {A.~P.}\ \bibnamefont {Bartok}},
  \bibinfo {author} {\bibfnamefont {G.}~\bibnamefont {Csanyi}}, \ and\ \bibinfo
  {author} {\bibfnamefont {M.}~\bibnamefont {Ceriotti}},\ }\href {\doibase
  10.1039/C6CP00415F} {\bibfield  {journal} {\bibinfo  {journal} {Phys. Chem.
  Chem. Phys.}\ }\textbf {\bibinfo {volume} {18}},\ \bibinfo {pages} {13754}
  (\bibinfo {year} {2016})}\BibitemShut {NoStop}%
\bibitem [{\citenamefont {Lee}\ \emph {et~al.}(2014)\citenamefont {Lee},
  \citenamefont {Urban}, \citenamefont {Li}, \citenamefont {Su}, \citenamefont
  {Hautier},\ and\ \citenamefont {Ceder}}]{sci343-2014-519}%
  \BibitemOpen
  \bibfield  {author} {\bibinfo {author} {\bibfnamefont {J.}~\bibnamefont
  {Lee}}, \bibinfo {author} {\bibfnamefont {A.}~\bibnamefont {Urban}}, \bibinfo
  {author} {\bibfnamefont {X.}~\bibnamefont {Li}}, \bibinfo {author}
  {\bibfnamefont {D.}~\bibnamefont {Su}}, \bibinfo {author} {\bibfnamefont
  {G.}~\bibnamefont {Hautier}}, \ and\ \bibinfo {author} {\bibfnamefont
  {G.}~\bibnamefont {Ceder}},\ }\href {\doibase 10.1126/science.1246432}
  {\bibfield  {journal} {\bibinfo  {journal} {Science}\ }\textbf {\bibinfo
  {volume} {343}},\ \bibinfo {pages} {519} (\bibinfo {year}
  {2014})}\BibitemShut {NoStop}%
\bibitem [{\citenamefont {Yabuuchi}\ \emph {et~al.}(2015)\citenamefont
  {Yabuuchi}, \citenamefont {Takeuchi}, \citenamefont {Nakayama}, \citenamefont
  {Shiiba}, \citenamefont {Ogawa}, \citenamefont {Nakayama}, \citenamefont
  {Ohta}, \citenamefont {Endo}, \citenamefont {Ozaki}, \citenamefont {Inamasu},
  \citenamefont {Sato},\ and\ \citenamefont {Komaba}}]{pnas112-2015-7650}%
  \BibitemOpen
  \bibfield  {author} {\bibinfo {author} {\bibfnamefont {N.}~\bibnamefont
  {Yabuuchi}}, \bibinfo {author} {\bibfnamefont {M.}~\bibnamefont {Takeuchi}},
  \bibinfo {author} {\bibfnamefont {M.}~\bibnamefont {Nakayama}}, \bibinfo
  {author} {\bibfnamefont {H.}~\bibnamefont {Shiiba}}, \bibinfo {author}
  {\bibfnamefont {M.}~\bibnamefont {Ogawa}}, \bibinfo {author} {\bibfnamefont
  {K.}~\bibnamefont {Nakayama}}, \bibinfo {author} {\bibfnamefont
  {T.}~\bibnamefont {Ohta}}, \bibinfo {author} {\bibfnamefont {D.}~\bibnamefont
  {Endo}}, \bibinfo {author} {\bibfnamefont {T.}~\bibnamefont {Ozaki}},
  \bibinfo {author} {\bibfnamefont {T.}~\bibnamefont {Inamasu}}, \bibinfo
  {author} {\bibfnamefont {K.}~\bibnamefont {Sato}}, \ and\ \bibinfo {author}
  {\bibfnamefont {S.}~\bibnamefont {Komaba}},\ }\href {\doibase
  10.1073/pnas.1504901112} {\bibfield  {journal} {\bibinfo  {journal} {Proc.
  Natl. Acad. Sci. USA}\ }\textbf {\bibinfo {volume} {112}},\ \bibinfo {pages}
  {7650–7655} (\bibinfo {year} {2015})}\BibitemShut {NoStop}%
\bibitem [{\citenamefont {Ropo}\ \emph
  {et~al.}(2016{\natexlab{a}})\citenamefont {Ropo}, \citenamefont {Schneider},
  \citenamefont {Baldauf},\ and\ \citenamefont {Blum}}]{sd3-2016-160009}%
  \BibitemOpen
  \bibfield  {author} {\bibinfo {author} {\bibfnamefont {M.}~\bibnamefont
  {Ropo}}, \bibinfo {author} {\bibfnamefont {M.}~\bibnamefont {Schneider}},
  \bibinfo {author} {\bibfnamefont {C.}~\bibnamefont {Baldauf}}, \ and\
  \bibinfo {author} {\bibfnamefont {V.}~\bibnamefont {Blum}},\ }\href {\doibase
  10.1038/sdata.2016.9} {\bibfield  {journal} {\bibinfo  {journal} {Sci. Data}\
  }\textbf {\bibinfo {volume} {3}},\ \bibinfo {pages} {160009} (\bibinfo {year}
  {2016}{\natexlab{a}})}\BibitemShut {NoStop}%
\bibitem [{\citenamefont {Ropo}\ \emph
  {et~al.}(2016{\natexlab{b}})\citenamefont {Ropo}, \citenamefont {Blum},\ and\
  \citenamefont {Baldauf}}]{sr6-2016-35772}%
  \BibitemOpen
  \bibfield  {author} {\bibinfo {author} {\bibfnamefont {M.}~\bibnamefont
  {Ropo}}, \bibinfo {author} {\bibfnamefont {V.}~\bibnamefont {Blum}}, \ and\
  \bibinfo {author} {\bibfnamefont {C.}~\bibnamefont {Baldauf}},\ }\href
  {\doibase 10.1038/srep35772} {\bibfield  {journal} {\bibinfo  {journal} {Sci.
  Rep.}\ }\textbf {\bibinfo {volume} {6}},\ \bibinfo {pages} {35772} (\bibinfo
  {year} {2016}{\natexlab{b}})}\BibitemShut {NoStop}%
\bibitem [{\citenamefont {Daw}\ and\ \citenamefont
  {Baskes}(1984)}]{prB29-1984-6443}%
  \BibitemOpen
  \bibfield  {author} {\bibinfo {author} {\bibfnamefont {M.~S.}\ \bibnamefont
  {Daw}}\ and\ \bibinfo {author} {\bibfnamefont {M.~I.}\ \bibnamefont
  {Baskes}},\ }\href {\doibase 10.1103/PhysRevB.29.6443} {\bibfield  {journal}
  {\bibinfo  {journal} {Phys. Rev. B}\ }\textbf {\bibinfo {volume} {29}},\
  \bibinfo {pages} {6443} (\bibinfo {year} {1984})}\BibitemShut {NoStop}%
\bibitem [{\citenamefont {Daw}\ \emph {et~al.}(1993)\citenamefont {Daw},
  \citenamefont {Foiles},\ and\ \citenamefont {Baskes}}]{msr9-1993-251}%
  \BibitemOpen
  \bibfield  {author} {\bibinfo {author} {\bibfnamefont {M.~S.}\ \bibnamefont
  {Daw}}, \bibinfo {author} {\bibfnamefont {S.~M.}\ \bibnamefont {Foiles}}, \
  and\ \bibinfo {author} {\bibfnamefont {M.~I.}\ \bibnamefont {Baskes}},\
  }\href@noop {} {\bibfield  {journal} {\bibinfo  {journal} {Mater. Sci. Rep.}\
  }\textbf {\bibinfo {volume} {9}},\ \bibinfo {pages} {251} (\bibinfo {year}
  {1993})}\BibitemShut {NoStop}%
\bibitem [{\citenamefont {Behler}(2011)}]{jcp134-2011-074106}%
  \BibitemOpen
  \bibfield  {author} {\bibinfo {author} {\bibfnamefont {J.}~\bibnamefont
  {Behler}},\ }\href {\doibase 10.1063/1.3553717} {\bibfield  {journal}
  {\bibinfo  {journal} {J. Chem. Phys.}\ }\textbf {\bibinfo {volume} {134}},\
  \bibinfo {pages} {074106} (\bibinfo {year} {2011})}\BibitemShut {NoStop}%
\bibitem [{\citenamefont {Bart\'ok}\ \emph {et~al.}(2013)\citenamefont
  {Bart\'ok}, \citenamefont {Kondor},\ and\ \citenamefont
  {Cs\'anyi}}]{prB87-2013-184115}%
  \BibitemOpen
  \bibfield  {author} {\bibinfo {author} {\bibfnamefont {A.~P.}\ \bibnamefont
  {Bart\'ok}}, \bibinfo {author} {\bibfnamefont {R.}~\bibnamefont {Kondor}}, \
  and\ \bibinfo {author} {\bibfnamefont {G.}~\bibnamefont {Cs\'anyi}},\ }\href
  {\doibase 10.1103/PhysRevB.87.184115} {\bibfield  {journal} {\bibinfo
  {journal} {Phys. Rev. B}\ }\textbf {\bibinfo {volume} {87}},\ \bibinfo
  {pages} {184115} (\bibinfo {year} {2013})}\BibitemShut {NoStop}%
\bibitem [{\citenamefont {Sadeghi}\ \emph {et~al.}(2013)\citenamefont
  {Sadeghi}, \citenamefont {Ghasemi}, \citenamefont {Schaefer}, \citenamefont
  {Mohr}, \citenamefont {Lill},\ and\ \citenamefont
  {Goedecker}}]{jcp139-2013-184118}%
  \BibitemOpen
  \bibfield  {author} {\bibinfo {author} {\bibfnamefont {A.}~\bibnamefont
  {Sadeghi}}, \bibinfo {author} {\bibfnamefont {S.~A.}\ \bibnamefont
  {Ghasemi}}, \bibinfo {author} {\bibfnamefont {B.}~\bibnamefont {Schaefer}},
  \bibinfo {author} {\bibfnamefont {S.}~\bibnamefont {Mohr}}, \bibinfo {author}
  {\bibfnamefont {M.~A.}\ \bibnamefont {Lill}}, \ and\ \bibinfo {author}
  {\bibfnamefont {S.}~\bibnamefont {Goedecker}},\ }\href {\doibase
  10.1063/1.4828704} {\bibfield  {journal} {\bibinfo  {journal} {J. Chem.
  Phys.}\ }\textbf {\bibinfo {volume} {139}},\ \bibinfo {pages} {184118}
  (\bibinfo {year} {2013})}\BibitemShut {NoStop}%
\bibitem [{\citenamefont {Sch\"utt}\ \emph {et~al.}(2014)\citenamefont
  {Sch\"utt}, \citenamefont {Glawe}, \citenamefont {Brockherde}, \citenamefont
  {Sanna}, \citenamefont {M\"uller},\ and\ \citenamefont
  {Gross}}]{prB89-2014-205118}%
  \BibitemOpen
  \bibfield  {author} {\bibinfo {author} {\bibfnamefont {K.~T.}\ \bibnamefont
  {Sch\"utt}}, \bibinfo {author} {\bibfnamefont {H.}~\bibnamefont {Glawe}},
  \bibinfo {author} {\bibfnamefont {F.}~\bibnamefont {Brockherde}}, \bibinfo
  {author} {\bibfnamefont {A.}~\bibnamefont {Sanna}}, \bibinfo {author}
  {\bibfnamefont {K.~R.}\ \bibnamefont {M\"uller}}, \ and\ \bibinfo {author}
  {\bibfnamefont {E.~K.~U.}\ \bibnamefont {Gross}},\ }\href {\doibase
  10.1103/PhysRevB.89.205118} {\bibfield  {journal} {\bibinfo  {journal} {Phys.
  Rev. B}\ }\textbf {\bibinfo {volume} {89}},\ \bibinfo {pages} {205118}
  (\bibinfo {year} {2014})}\BibitemShut {NoStop}%
\bibitem [{\citenamefont {Behler}(2015)}]{ijqc115-2015-1032}%
  \BibitemOpen
  \bibfield  {author} {\bibinfo {author} {\bibfnamefont {J.}~\bibnamefont
  {Behler}},\ }\href {\doibase 10.1002/qua.24890} {\bibfield  {journal}
  {\bibinfo  {journal} {Int. J. Quantum Chem.}\ }\textbf {\bibinfo {volume}
  {115}},\ \bibinfo {pages} {1032} (\bibinfo {year} {2015})}\BibitemShut
  {NoStop}%
\bibitem [{\citenamefont {Faber}\ \emph {et~al.}(2015)\citenamefont {Faber},
  \citenamefont {Lindmaa}, \citenamefont {von Lilienfeld},\ and\ \citenamefont
  {Armiento}}]{ijqc115-2015-1094}%
  \BibitemOpen
  \bibfield  {author} {\bibinfo {author} {\bibfnamefont {F.}~\bibnamefont
  {Faber}}, \bibinfo {author} {\bibfnamefont {A.}~\bibnamefont {Lindmaa}},
  \bibinfo {author} {\bibfnamefont {O.~A.}\ \bibnamefont {von Lilienfeld}}, \
  and\ \bibinfo {author} {\bibfnamefont {R.}~\bibnamefont {Armiento}},\ }\href
  {\doibase 10.1002/qua.24917} {\bibfield  {journal} {\bibinfo  {journal} {Int.
  J. Quantum Chem.}\ }\textbf {\bibinfo {volume} {115}},\ \bibinfo {pages}
  {1094} (\bibinfo {year} {2015})}\BibitemShut {NoStop}%
\bibitem [{\citenamefont {von Lilienfeld}\ \emph {et~al.}(2015)\citenamefont
  {von Lilienfeld}, \citenamefont {Ramakrishnan}, \citenamefont {Rupp},\ and\
  \citenamefont {Knoll}}]{ijqc115-2015-1084}%
  \BibitemOpen
  \bibfield  {author} {\bibinfo {author} {\bibfnamefont {O.~A.}\ \bibnamefont
  {von Lilienfeld}}, \bibinfo {author} {\bibfnamefont {R.}~\bibnamefont
  {Ramakrishnan}}, \bibinfo {author} {\bibfnamefont {M.}~\bibnamefont {Rupp}},
  \ and\ \bibinfo {author} {\bibfnamefont {A.}~\bibnamefont {Knoll}},\ }\href
  {\doibase 10.1002/qua.24912} {\bibfield  {journal} {\bibinfo  {journal} {Int.
  J. Quantum Chem.}\ }\textbf {\bibinfo {volume} {115}},\ \bibinfo {pages}
  {1084} (\bibinfo {year} {2015})}\BibitemShut {NoStop}%
\bibitem [{\citenamefont {Bart\'{o}k}\ and\ \citenamefont
  {Cs\'{a}nyi}(2015)}]{ijqc115-2015-1051}%
  \BibitemOpen
  \bibfield  {author} {\bibinfo {author} {\bibfnamefont {A.~P.}\ \bibnamefont
  {Bart\'{o}k}}\ and\ \bibinfo {author} {\bibfnamefont {G.}~\bibnamefont
  {Cs\'{a}nyi}},\ }\href {\doibase 10.1002/qua.24927} {\bibfield  {journal}
  {\bibinfo  {journal} {Int. J. Quantum Chem.}\ }\textbf {\bibinfo {volume}
  {115}},\ \bibinfo {pages} {1051} (\bibinfo {year} {2015})}\BibitemShut
  {NoStop}%
\bibitem [{\citenamefont {Artrith}\ and\ \citenamefont
  {Urban}(2016)}]{cms114-2016-135}%
  \BibitemOpen
  \bibfield  {author} {\bibinfo {author} {\bibfnamefont {N.}~\bibnamefont
  {Artrith}}\ and\ \bibinfo {author} {\bibfnamefont {A.}~\bibnamefont
  {Urban}},\ }\href {\doibase 10.1016/j.commatsci.2015.11.047} {\bibfield
  {journal} {\bibinfo  {journal} {Comput. Mater. Sci.}\ }\textbf {\bibinfo
  {volume} {114}},\ \bibinfo {pages} {135} (\bibinfo {year}
  {2016})}\BibitemShut {NoStop}%
\bibitem [{\citenamefont {Hornik}(1991)}]{nn4-1991-251}%
  \BibitemOpen
  \bibfield  {author} {\bibinfo {author} {\bibfnamefont {K.}~\bibnamefont
  {Hornik}},\ }\href {\doibase 10.1016/0893-6080(91)90009-t} {\bibfield
  {journal} {\bibinfo  {journal} {Neural Networks}\ }\textbf {\bibinfo {volume}
  {4}},\ \bibinfo {pages} {251} (\bibinfo {year} {1991})}\BibitemShut {NoStop}%
\bibitem [{\citenamefont {Liechtenstein}\ \emph {et~al.}(1995)\citenamefont
  {Liechtenstein}, \citenamefont {Anisimov},\ and\ \citenamefont
  {Zaanen}}]{prB52-1995-R5467}%
  \BibitemOpen
  \bibfield  {author} {\bibinfo {author} {\bibfnamefont {A.~I.}\ \bibnamefont
  {Liechtenstein}}, \bibinfo {author} {\bibfnamefont {V.~I.}\ \bibnamefont
  {Anisimov}}, \ and\ \bibinfo {author} {\bibfnamefont {J.}~\bibnamefont
  {Zaanen}},\ }\href {\doibase 10.1103/PhysRevB.52.R5467} {\bibfield  {journal}
  {\bibinfo  {journal} {Phys. Rev. B}\ }\textbf {\bibinfo {volume} {52}},\
  \bibinfo {pages} {R5467} (\bibinfo {year} {1995})}\BibitemShut {NoStop}%
\bibitem [{\citenamefont {Anisimov}\ \emph {et~al.}(1997)\citenamefont
  {Anisimov}, \citenamefont {Aryasetiawan},\ and\ \citenamefont
  {Lichtenstein}}]{jpcm9-1997-767}%
  \BibitemOpen
  \bibfield  {author} {\bibinfo {author} {\bibfnamefont {V.~I.}\ \bibnamefont
  {Anisimov}}, \bibinfo {author} {\bibfnamefont {F.}~\bibnamefont
  {Aryasetiawan}}, \ and\ \bibinfo {author} {\bibfnamefont {A.~I.}\
  \bibnamefont {Lichtenstein}},\ }\href@noop {} {\bibfield  {journal} {\bibinfo
   {journal} {J. Phys.: Condens. Matter}\ }\textbf {\bibinfo {volume} {9}},\
  \bibinfo {pages} {767} (\bibinfo {year} {1997})}\BibitemShut {NoStop}%
\bibitem [{\citenamefont {Jain}\ \emph {et~al.}(2011)\citenamefont {Jain},
  \citenamefont {Hautier}, \citenamefont {Moore}, \citenamefont {Ong},
  \citenamefont {Fischer}, \citenamefont {Mueller}, \citenamefont {Persson},\
  and\ \citenamefont {Ceder}}]{cms50-2011-2295}%
  \BibitemOpen
  \bibfield  {author} {\bibinfo {author} {\bibfnamefont {A.}~\bibnamefont
  {Jain}}, \bibinfo {author} {\bibfnamefont {G.}~\bibnamefont {Hautier}},
  \bibinfo {author} {\bibfnamefont {C.~J.}\ \bibnamefont {Moore}}, \bibinfo
  {author} {\bibfnamefont {S.~P.}\ \bibnamefont {Ong}}, \bibinfo {author}
  {\bibfnamefont {C.~C.}\ \bibnamefont {Fischer}}, \bibinfo {author}
  {\bibfnamefont {T.}~\bibnamefont {Mueller}}, \bibinfo {author} {\bibfnamefont
  {K.~A.}\ \bibnamefont {Persson}}, \ and\ \bibinfo {author} {\bibfnamefont
  {G.}~\bibnamefont {Ceder}},\ }\href@noop {} {\bibfield  {journal} {\bibinfo
  {journal} {Comput. Mater. Sci.}\ }\textbf {\bibinfo {volume} {50}},\ \bibinfo
  {pages} {2295} (\bibinfo {year} {2011})}\BibitemShut {NoStop}%
\bibitem [{\citenamefont {Perdew}\ \emph {et~al.}(1996)\citenamefont {Perdew},
  \citenamefont {Burke},\ and\ \citenamefont {Ernzerhof}}]{prl77-96-3865}%
  \BibitemOpen
  \bibfield  {author} {\bibinfo {author} {\bibfnamefont {J.}~\bibnamefont
  {Perdew}}, \bibinfo {author} {\bibfnamefont {K.}~\bibnamefont {Burke}}, \
  and\ \bibinfo {author} {\bibfnamefont {M.}~\bibnamefont {Ernzerhof}},\ }\href
  {\doibase 10.1103/PhysRevLett.77.3865} {\bibfield  {journal} {\bibinfo
  {journal} {Phys. Rev. Lett.}\ }\textbf {\bibinfo {volume} {77}},\ \bibinfo
  {pages} {3865} (\bibinfo {year} {1996})}\BibitemShut {NoStop}%
\bibitem [{\citenamefont {Bl\"ochl}(1994)}]{prB50-1994-17953}%
  \BibitemOpen
  \bibfield  {author} {\bibinfo {author} {\bibfnamefont {P.~E.}\ \bibnamefont
  {Bl\"ochl}},\ }\href {\doibase 10.1103/PhysRevB.50.17953} {\bibfield
  {journal} {\bibinfo  {journal} {Phys. Rev. B}\ }\textbf {\bibinfo {volume}
  {50}},\ \bibinfo {pages} {17953} (\bibinfo {year} {1994})}\BibitemShut
  {NoStop}%
\bibitem [{\citenamefont {Kresse}\ and\ \citenamefont
  {Furthm\"uller}(1996{\natexlab{a}})}]{prB54-1996-11169}%
  \BibitemOpen
  \bibfield  {author} {\bibinfo {author} {\bibfnamefont {G.}~\bibnamefont
  {Kresse}}\ and\ \bibinfo {author} {\bibfnamefont {J.}~\bibnamefont
  {Furthm\"uller}},\ }\href {\doibase 10.1103/PhysRevB.54.11169} {\bibfield
  {journal} {\bibinfo  {journal} {Phys. Rev. B}\ }\textbf {\bibinfo {volume}
  {54}},\ \bibinfo {pages} {11169} (\bibinfo {year}
  {1996}{\natexlab{a}})}\BibitemShut {NoStop}%
\bibitem [{\citenamefont {Kresse}\ and\ \citenamefont
  {Furthm\"uller}(1996{\natexlab{b}})}]{cms6-1996-15}%
  \BibitemOpen
  \bibfield  {author} {\bibinfo {author} {\bibfnamefont {G.}~\bibnamefont
  {Kresse}}\ and\ \bibinfo {author} {\bibfnamefont {J.}~\bibnamefont
  {Furthm\"uller}},\ }\href {\doibase 10.1016/0927-0256(96)00008-0} {\bibfield
  {journal} {\bibinfo  {journal} {Comput. Mater. Sci.}\ }\textbf {\bibinfo
  {volume} {6}},\ \bibinfo {pages} {15} (\bibinfo {year}
  {1996}{\natexlab{b}})}\BibitemShut {NoStop}%
\bibitem [{\citenamefont {Ong}\ \emph {et~al.}(2013)\citenamefont {Ong},
  \citenamefont {Richards}, \citenamefont {Jain}, \citenamefont {Hautier},
  \citenamefont {Kocher}, \citenamefont {Cholia}, \citenamefont {Gunter},
  \citenamefont {Chevrier}, \citenamefont {Persson},\ and\ \citenamefont
  {Ceder}}]{cms68-2013-314}%
  \BibitemOpen
  \bibfield  {author} {\bibinfo {author} {\bibfnamefont {S.~P.}\ \bibnamefont
  {Ong}}, \bibinfo {author} {\bibfnamefont {W.~D.}\ \bibnamefont {Richards}},
  \bibinfo {author} {\bibfnamefont {A.}~\bibnamefont {Jain}}, \bibinfo {author}
  {\bibfnamefont {G.}~\bibnamefont {Hautier}}, \bibinfo {author} {\bibfnamefont
  {M.}~\bibnamefont {Kocher}}, \bibinfo {author} {\bibfnamefont
  {S.}~\bibnamefont {Cholia}}, \bibinfo {author} {\bibfnamefont
  {D.}~\bibnamefont {Gunter}}, \bibinfo {author} {\bibfnamefont {V.~L.}\
  \bibnamefont {Chevrier}}, \bibinfo {author} {\bibfnamefont {K.~A.}\
  \bibnamefont {Persson}}, \ and\ \bibinfo {author} {\bibfnamefont
  {G.}~\bibnamefont {Ceder}},\ }\href {\doibase
  10.1016/j.commatsci.2012.10.028} {\bibfield  {journal} {\bibinfo  {journal}
  {Comput. Mater. Sci.}\ }\textbf {\bibinfo {volume} {68}},\ \bibinfo {pages}
  {314} (\bibinfo {year} {2013})}\BibitemShut {NoStop}%
\bibitem [{\citenamefont {Byrd}\ \emph {et~al.}(1995)\citenamefont {Byrd},
  \citenamefont {Lu}, \citenamefont {Nocedal},\ and\ \citenamefont
  {Zhu}}]{siam16-1995-1190}%
  \BibitemOpen
  \bibfield  {author} {\bibinfo {author} {\bibfnamefont {R.}~\bibnamefont
  {Byrd}}, \bibinfo {author} {\bibfnamefont {P.}~\bibnamefont {Lu}}, \bibinfo
  {author} {\bibfnamefont {J.}~\bibnamefont {Nocedal}}, \ and\ \bibinfo
  {author} {\bibfnamefont {C.}~\bibnamefont {Zhu}},\ }\href {\doibase
  10.1137/0916069} {\bibfield  {journal} {\bibinfo  {journal} {SIAM J. Sci.
  Comput.}\ }\textbf {\bibinfo {volume} {16}},\ \bibinfo {pages} {1190}
  (\bibinfo {year} {1995})}\BibitemShut {NoStop}%
\bibitem [{\citenamefont {Zhu}\ \emph {et~al.}(1997)\citenamefont {Zhu},
  \citenamefont {Byrd}, \citenamefont {Lu},\ and\ \citenamefont
  {Nocedal}}]{toms23-1997-550}%
  \BibitemOpen
  \bibfield  {author} {\bibinfo {author} {\bibfnamefont {C.}~\bibnamefont
  {Zhu}}, \bibinfo {author} {\bibfnamefont {R.~H.}\ \bibnamefont {Byrd}},
  \bibinfo {author} {\bibfnamefont {P.}~\bibnamefont {Lu}}, \ and\ \bibinfo
  {author} {\bibfnamefont {J.}~\bibnamefont {Nocedal}},\ }\href {\doibase
  10.1145/279232.279236} {\bibfield  {journal} {\bibinfo  {journal} {ACM T.
  Math Software}\ }\textbf {\bibinfo {volume} {23}},\ \bibinfo {pages}
  {550–560} (\bibinfo {year} {1997})}\BibitemShut {NoStop}%
\bibitem [{\citenamefont {Tkatchenko}\ and\ \citenamefont
  {Scheffler}(2009)}]{prl102-09-073005}%
  \BibitemOpen
  \bibfield  {author} {\bibinfo {author} {\bibfnamefont {A.}~\bibnamefont
  {Tkatchenko}}\ and\ \bibinfo {author} {\bibfnamefont {M.}~\bibnamefont
  {Scheffler}},\ }\href {\doibase 10.1103/PhysRevLett.102.073005} {\bibfield
  {journal} {\bibinfo  {journal} {Phys. Rev. Lett.}\ }\textbf {\bibinfo
  {volume} {102}},\ \bibinfo {pages} {073005} (\bibinfo {year}
  {2009})}\BibitemShut {NoStop}%
\bibitem [{\citenamefont {Blum}\ \emph {et~al.}(2009)\citenamefont {Blum},
  \citenamefont {Gehrke}, \citenamefont {Hanke}, \citenamefont {Havu},
  \citenamefont {Havu}, \citenamefont {Ren}, \citenamefont {Reuter},\ and\
  \citenamefont {Scheffler}}]{cpc180-2009-2175}%
  \BibitemOpen
  \bibfield  {author} {\bibinfo {author} {\bibfnamefont {V.}~\bibnamefont
  {Blum}}, \bibinfo {author} {\bibfnamefont {R.}~\bibnamefont {Gehrke}},
  \bibinfo {author} {\bibfnamefont {F.}~\bibnamefont {Hanke}}, \bibinfo
  {author} {\bibfnamefont {P.}~\bibnamefont {Havu}}, \bibinfo {author}
  {\bibfnamefont {V.}~\bibnamefont {Havu}}, \bibinfo {author} {\bibfnamefont
  {X.}~\bibnamefont {Ren}}, \bibinfo {author} {\bibfnamefont {K.}~\bibnamefont
  {Reuter}}, \ and\ \bibinfo {author} {\bibfnamefont {M.}~\bibnamefont
  {Scheffler}},\ }\href {\doibase 10.1016/j.cpc.2009.06.022} {\bibfield
  {journal} {\bibinfo  {journal} {Comput. Phys. Commun.}\ }\textbf {\bibinfo
  {volume} {180}},\ \bibinfo {pages} {2175} (\bibinfo {year}
  {2009})}\BibitemShut {NoStop}%
\bibitem [{\citenamefont {Piana}\ \emph {et~al.}(2014)\citenamefont {Piana},
  \citenamefont {Klepeis},\ and\ \citenamefont {Shaw}}]{coisb24-2014-98}%
  \BibitemOpen
  \bibfield  {author} {\bibinfo {author} {\bibfnamefont {S.}~\bibnamefont
  {Piana}}, \bibinfo {author} {\bibfnamefont {J.~L.}\ \bibnamefont {Klepeis}},
  \ and\ \bibinfo {author} {\bibfnamefont {D.~E.}\ \bibnamefont {Shaw}},\
  }\href {\doibase 10.1016/j.sbi.2013.12.006} {\bibfield  {journal} {\bibinfo
  {journal} {Curr. Opin. Struct. Biol.}\ }\textbf {\bibinfo {volume} {24}},\
  \bibinfo {pages} {98} (\bibinfo {year} {2014})}\BibitemShut {NoStop}%
\bibitem [{\citenamefont {Sanchez}\ \emph {et~al.}(1984)\citenamefont
  {Sanchez}, \citenamefont {Ducastelle},\ and\ \citenamefont
  {Gratias}}]{pa128-1984-334}%
  \BibitemOpen
  \bibfield  {author} {\bibinfo {author} {\bibfnamefont {J.}~\bibnamefont
  {Sanchez}}, \bibinfo {author} {\bibfnamefont {F.}~\bibnamefont {Ducastelle}},
  \ and\ \bibinfo {author} {\bibfnamefont {D.}~\bibnamefont {Gratias}},\
  }\href@noop {} {\bibfield  {journal} {\bibinfo  {journal} {Physica A}\
  }\textbf {\bibinfo {volume} {128}},\ \bibinfo {pages} {334} (\bibinfo {year}
  {1984})}\BibitemShut {NoStop}%
\bibitem [{\citenamefont {Fontaine}(1994)}]{ssp47-1994-33}%
  \BibitemOpen
  \bibfield  {author} {\bibinfo {author} {\bibfnamefont {D.~D.}\ \bibnamefont
  {Fontaine}},\ }\href {\doibase 10.1016/s0081-1947(08)60639-6} {\bibfield
  {journal} {\bibinfo  {journal} {Solid State Phys.}\ }\textbf {\bibinfo
  {volume} {47}},\ \bibinfo {pages} {33–176} (\bibinfo {year}
  {1994})}\BibitemShut {NoStop}%
\bibitem [{\citenamefont {Ceder}(1993)}]{cms1-1993-144}%
  \BibitemOpen
  \bibfield  {author} {\bibinfo {author} {\bibfnamefont {G.}~\bibnamefont
  {Ceder}},\ }\href@noop {} {\bibfield  {journal} {\bibinfo  {journal} {Comput.
  Mater. Sci.}\ }\textbf {\bibinfo {volume} {1}},\ \bibinfo {pages} {144}
  (\bibinfo {year} {1993})}\BibitemShut {NoStop}%
\bibitem [{\citenamefont {Thompson}\ \emph {et~al.}(2014)\citenamefont
  {Thompson}, \citenamefont {Swiler}, \citenamefont {Trott}, \citenamefont
  {Foiles},\ and\ \citenamefont {Tucker}}]{jocp-2014-SNAP}%
  \BibitemOpen
  \bibfield  {author} {\bibinfo {author} {\bibfnamefont {A.}~\bibnamefont
  {Thompson}}, \bibinfo {author} {\bibfnamefont {L.}~\bibnamefont {Swiler}},
  \bibinfo {author} {\bibfnamefont {C.}~\bibnamefont {Trott}}, \bibinfo
  {author} {\bibfnamefont {S.}~\bibnamefont {Foiles}}, \ and\ \bibinfo {author}
  {\bibfnamefont {G.}~\bibnamefont {Tucker}},\ }\href {\doibase
  10.1016/j.jcp.2014.12.018} {\bibfield  {journal} {\bibinfo  {journal} {J.
  Comput. Phys.}\ }\textbf {\bibinfo {volume} {285}},\ \bibinfo {pages} {316}
  (\bibinfo {year} {2014})}\BibitemShut {NoStop}%
\bibitem [{\citenamefont {Rogati}\ and\ \citenamefont
  {Yang}(2002)}]{RogatiYang2002}%
  \BibitemOpen
  \bibfield  {author} {\bibinfo {author} {\bibfnamefont {M.}~\bibnamefont
  {Rogati}}\ and\ \bibinfo {author} {\bibfnamefont {Y.}~\bibnamefont {Yang}},\
  }in\ \href {\doibase 10.1145/584792.584911} {\emph {\bibinfo {booktitle}
  {Proceedings of the eleventh international conference on Information and
  knowledge management - {CIKM} '02}}}\ (\bibinfo  {publisher} {{ACM} Press},\
  \bibinfo {year} {2002})\BibitemShut {NoStop}%
\bibitem [{\citenamefont {Urban}\ \emph {et~al.}(2016)\citenamefont {Urban},
  \citenamefont {Matts}, \citenamefont {Abdellahi},\ and\ \citenamefont
  {Ceder}}]{aem-2016-1600488}%
  \BibitemOpen
  \bibfield  {author} {\bibinfo {author} {\bibfnamefont {A.}~\bibnamefont
  {Urban}}, \bibinfo {author} {\bibfnamefont {I.}~\bibnamefont {Matts}},
  \bibinfo {author} {\bibfnamefont {A.}~\bibnamefont {Abdellahi}}, \ and\
  \bibinfo {author} {\bibfnamefont {G.}~\bibnamefont {Ceder}},\ }\href
  {\doibase 10.1002/aenm.201600488} {\bibfield  {journal} {\bibinfo  {journal}
  {Adv. Energy Mater.}\ }\textbf {\bibinfo {volume} {6}},\ \bibinfo {pages}
  {1600488} (\bibinfo {year} {2016})}\BibitemShut {NoStop}%
\bibitem [{\citenamefont {Hart}\ and\ \citenamefont
  {Forcade}(2008)}]{prB77-2008-224115}%
  \BibitemOpen
  \bibfield  {author} {\bibinfo {author} {\bibfnamefont {G.~L.~W.}\
  \bibnamefont {Hart}}\ and\ \bibinfo {author} {\bibfnamefont {R.~W.}\
  \bibnamefont {Forcade}},\ }\href {\doibase 10.1103/PhysRevB.77.224115}
  {\bibfield  {journal} {\bibinfo  {journal} {Phys. Rev. B}\ }\textbf {\bibinfo
  {volume} {77}},\ \bibinfo {pages} {224115} (\bibinfo {year}
  {2008})}\BibitemShut {NoStop}%
\bibitem [{\citenamefont {Hart}\ and\ \citenamefont
  {Forcade}(2009)}]{prB80-2009-014120}%
  \BibitemOpen
  \bibfield  {author} {\bibinfo {author} {\bibfnamefont {G.~L.~W.}\
  \bibnamefont {Hart}}\ and\ \bibinfo {author} {\bibfnamefont {R.~W.}\
  \bibnamefont {Forcade}},\ }\href {\doibase 10.1103/PhysRevB.80.014120}
  {\bibfield  {journal} {\bibinfo  {journal} {Phys. Rev. B}\ }\textbf {\bibinfo
  {volume} {80}},\ \bibinfo {pages} {014120} (\bibinfo {year}
  {2009})}\BibitemShut {NoStop}%
\bibitem [{\citenamefont {Hart}\ \emph {et~al.}(2012)\citenamefont {Hart},
  \citenamefont {Nelson},\ and\ \citenamefont {Forcade}}]{cms59-2012-101}%
  \BibitemOpen
  \bibfield  {author} {\bibinfo {author} {\bibfnamefont {G.~L.}\ \bibnamefont
  {Hart}}, \bibinfo {author} {\bibfnamefont {L.~J.}\ \bibnamefont {Nelson}}, \
  and\ \bibinfo {author} {\bibfnamefont {R.~W.}\ \bibnamefont {Forcade}},\
  }\href {\doibase 10.1016/j.commatsci.2012.02.015} {\bibfield  {journal}
  {\bibinfo  {journal} {Comput. Mater. Sci.}\ }\textbf {\bibinfo {volume}
  {59}},\ \bibinfo {pages} {101} (\bibinfo {year} {2012})}\BibitemShut
  {NoStop}%
\bibitem [{\citenamefont {Artrith}\ \emph {et~al.}(2013)\citenamefont
  {Artrith}, \citenamefont {Hiller},\ and\ \citenamefont
  {Behler}}]{pssb250-2013-1191}%
  \BibitemOpen
  \bibfield  {author} {\bibinfo {author} {\bibfnamefont {N.}~\bibnamefont
  {Artrith}}, \bibinfo {author} {\bibfnamefont {B.}~\bibnamefont {Hiller}}, \
  and\ \bibinfo {author} {\bibfnamefont {J.}~\bibnamefont {Behler}},\ }\href
  {\doibase 10.1002/pssb.201248370} {\bibfield  {journal} {\bibinfo  {journal}
  {physica status solidi (b)}\ }\textbf {\bibinfo {volume} {250}},\ \bibinfo
  {pages} {1191} (\bibinfo {year} {2013})}\BibitemShut {NoStop}%
\end{thebibliography}%


\newpage
\appendix
\renewcommand{\thefigure}{S\arabic{figure}}
\renewcommand{\thetable}{S\arabic{table}}
\setcounter{figure}{0}
\setcounter{table}{0}

\begin{widetext}
\begin{figure*}
  \centering
  \includegraphics[width=0.9\textwidth]{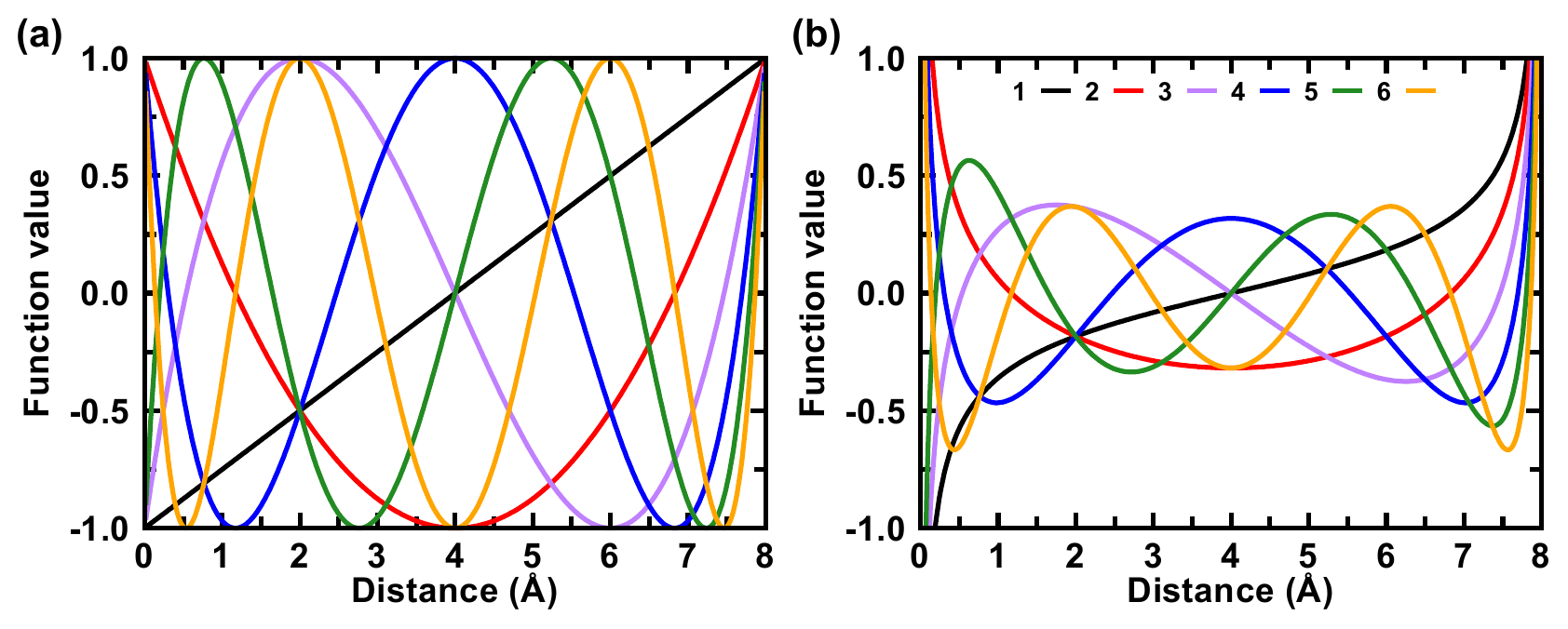}
  \caption{\label{fig:chebyshev}%
    \textbf{(a)}~Dual basis functions of Eq.~\eqref{eq:phi'} (Chebyshev
    polynomials) up to order $\alpha=6$ for a catoff radius
    $R_{\textup{c}}=8.0$~\AA{}.  The polynomial of order $\alpha=0$ is
    constant~1 and not shown.  \textbf{(b)}~The corresponding basis
    functions of Eq.~\eqref{eq:phi} (only needed for the reconstruction
    of the RDF or ADF).}
  \vspace{-\baselineskip}
\end{figure*}
\end{widetext}

\begin{figure*}
  \centering
  \includegraphics[width=0.9\textwidth]{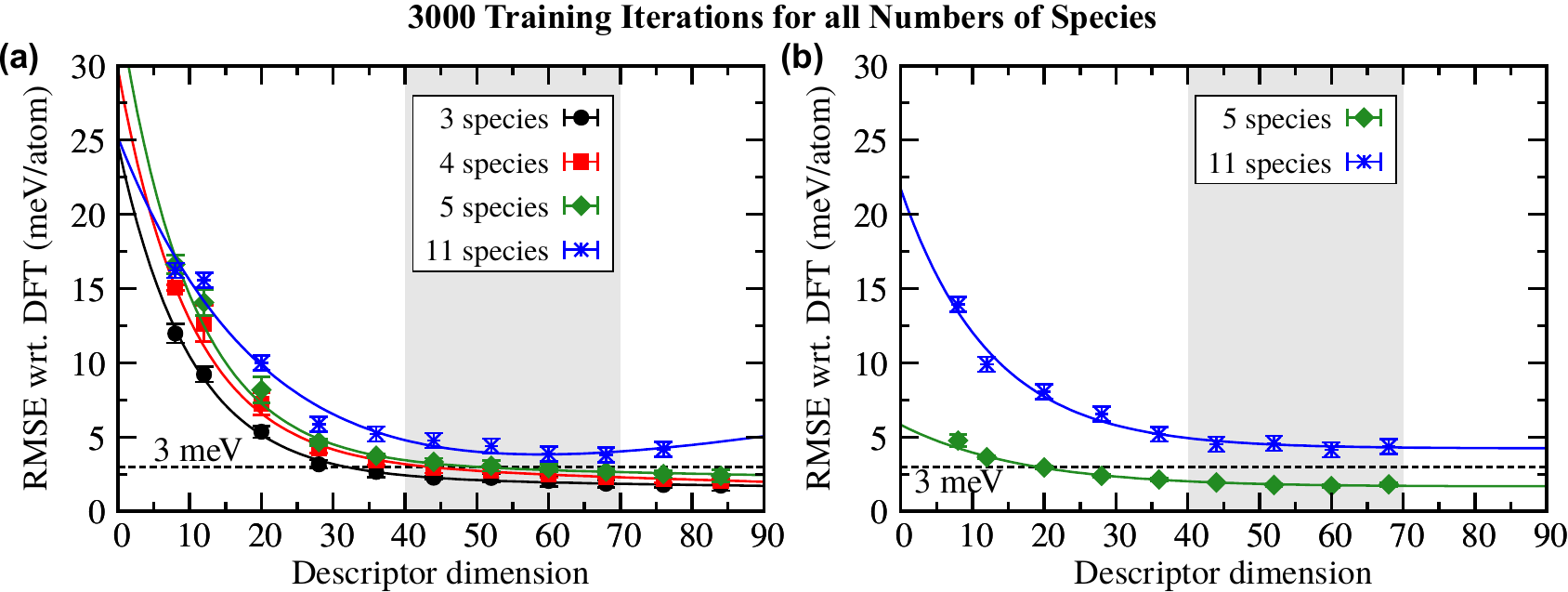}
  \caption{\label{fig:complexity-3000}%
    Precision of artificial neural network (ANN) potentials as function
    of the dimension of the descriptor used to represent the local
    atomic environment.
    The root mean squared error (RMSE) after 3,000~training iterations
    is shown.
    After this number of iterations, the RMSE has not yet converged for
    the 11-species systems and is increasing with descriptor size.  }
\end{figure*}

\section{The $\textup{Li}M\textup{O}_{2}$ data set}
\label{sec:LiMO2}


Starting point for the generation of the \ce{Li$M$O2} data set of this
work were the enumerated lithium transition-metal (TM) oxide
configurations of reference~\onlinecite{aem-2016-1600488}.
The data set with 3~chemical species (Li, Ti, and O) comprised a total
of 7,338~structures including the \ce{TiO2} structures from
reference~\onlinecite{cms114-2016-135} and additional \ce{LiTiO2}
configurations that were systematically enumerated based on the rocksalt
structure (A sites = Li and Ti, B sites = O) up to cell sizes containing
8~cation sites using the approach by Hart
et~al.~\cite{prB77-2008-224115, prB80-2009-014120, cms59-2012-101}.
For the data set with 4~chemical species (Li, Ni, Ti, and O), 1,343
atomic configurations with compositions \ce{LiNiO2} and \ce{Li2NiTiO4}
were additionally generated using the same enumeration methodology
(giving a total of 8,681 configurations).
Further, 1,494 atomic configurations with compositions \ce{LiMnO2} and
\ce{Li2NiMnO4} were added for the set with 5~species (Li, Ti, Mn, Ni, O)
to a total of 10,175 configurations.
Finally, for 11~chemical species, random atomic configurations with
composition \ce{Li9$M$9O18} with $M$\,=\,Sc, Ti, V, Cr, Mn, Fe, Co, Ni,
Cu for all 24,310~possible compositions were generated, and 5,872
randomly selected configurations were included in the reference data
set.

The complete reference data set for 11 chemical species comprises a
total of 16,047 atomic configurations.

\section{Artificial neural network potentials}
\label{sec:ANN}

Together with the dimension of the descriptor discussed in the main
text, the architecture of an artificial neural network (ANN) determines
the model complexity.
For the feedforward ANNs used in the present work, the architecture is
given by the number of hidden layers and the number of nodes per layer
employed by the ANN (see also reference~\cite{cms114-2016-135} for a
detailed introduction).
To facilitate comparison of the different structure and composition
spaces on equal footing, we generally used a $N$-15-15-1 ANN
architecture, i.e., an architecture with two hidden layeres containing
each 15~nodes independent of the descriptor dimension $N$.

The atomic energy network (aenet) software~\cite{cms114-2016-135} was
employed for the training of the ANN potentials with the limited-memory
BFGS method~\cite{siam16-1995-1190, toms23-1997-550}.

\section{Scaling behavior of existing local descriptors}
\label{sec:scaling}

The complexity of the Behler--Parrinello (BP) descriptor and the
cluster-expansion basis scales at least quadratically with the number of
chemical species.

As noted in the main manuscript, separate MLPs for each chemical species
are constructed. For each of these MLPs, the descriptor also scales with
the number of species:

For example, the angular symmetry functions for an ANN potential for a
single species \textbf{A} describe the interactions of the central atom
with two atoms of type \textbf{A} (\textbf{A-A}).  For two species
\textbf{A} and \textbf{B}, three interactions occur (\textbf{A-A},
\textbf{A-B}, and \textbf{B-B}), and for three species \textbf{A},
\textbf{B}, \textbf{C}, there are six (\textbf{A-A}, \textbf{A-B},
\textbf{A-C}, \textbf{B-B}, \textbf{B-C}, \textbf{C-C}).
In general, for $N$ species the number of interactions is $N(N+1)/2$,
i.e., the scaling is quadratic in the number of species.
Further details about the symmetry function set up for multicomponent
systems using the Behler-Parrinello approach along with actual
parameters can also be found in
reference~\onlinecite{pssb250-2013-1191}.

When Ising-like pseudo spin variables are used to describe compositions,
as for example in the cluster expansion (CE) method, an analogous
scaling with the number of species occurs.  The number of CE basis
functions scales quadractically with the number of species when only
pair clusters are considered.  Generally, the scaling is on the order of
the highest included n-body interaction, i.e., cubic for triplets, 4th
order for quadruplet interactions, and so on.  Mathematically, this
relationship was worked out in reference~\onlinecite{pa128-1984-334}.

The multi-component implementation of the smooth overlap of atomic
positions (SOAP) approach, is laid out in section 2.3~of
reference~\onlinecite{pccp18-2016-13754}.  The scaling is also
quadratic, as it involves partial power spectra for each pair of
species.

\enlargethispage{3\baselineskip}
\section{Derivation of the expansion coefficients}
\label{sec:coeff}

The expansion coefficient $c^{(2)}_{\alpha}$ of the basis set expansion
of the radial distribution function (RDF)
\begin{align}
  \mathrm{RDF}_{i}(r)
  = \sum_{\alpha} c^{(2)}_{\alpha} \phi_{\alpha}(r)
  \quad ,
  \label{eq:RDF-expansion}
\end{align}
where
$\int\bar{\phi}_{\beta}(r)\phi_{\alpha}(r)\,\mathrm{d}r=\delta_{\beta\alpha}$
and $\{\bar{\phi}_{\alpha}\}$ is the orthogonal \emph{dual} basis to
$\{\phi_{\alpha}\}$, is given by
\begin{align}
  c^{(2)}_{\alpha}
  = \!\int_{0}^{R_{\textup{c}}}\!
    \bar{\phi}_{\alpha}(r)\, \mathrm{RDF}_{i}(r) \,\mathrm{d}r
  \label{eq:c^(2)}
  \quad .
\end{align}
Note that the RDF as defined in Eq.~(4) of the main manuscript
\begin{align}
  \mathrm{RDF}_{i}(r)
  &= \sum_{\mathclap{\;\;\mathbf{R}_{j}\in\,\sigma_{i}^{R_{\textup{c}}}}}
    \delta(r - R_{ij}) \,f_{\textup{c}}(R_{ij}) \,w_{t_{j}}
  \label{eq:RDF}
\end{align}
is only different from $0$ for $0\leq{}r\leq{}R_{\textup{c}}$, so that
the integral in Eq.~\eqref{eq:c^(2)} can be replaced by an integral over
the entire space.
Inserting the expression of the RDF Eq.~\eqref{eq:RDF} into the
Eq.~\eqref{eq:c^(2)} yields
\begin{align}
  c^{(2)}_{\alpha}
  &= \sum_{\mathclap{\;\;\mathbf{R}_{j}\in\,\sigma_{i}^{R_{\textup{c}}}}}\quad
      \int\! \bar{\phi}_{\alpha}(r)\, \delta(r - R_{ij})\,f_{\textup{c}}(R_{ij}) \,w_{t_{j}} \,\mathrm{d}r
    \nonumber \\
  &= \sum_{\mathclap{\;\;\mathbf{R}_{j}\in\,\sigma_{i}^{R_{\textup{c}}}}}
     \bar{\phi}_{\alpha}(R_{ij}) \,f_{\textup{c}}(R_{ij}) \,w_{t_{j}}
  \quad ,
  \label{eq:coeff-(2)}
\end{align}
which is the expression given in Eq.~(6) of the main manuscript.

The derivation of the coefficients $c^{(3)}_{\alpha}$ of the angular
expansion is completely analogous.

\enlargethispage{3\baselineskip}
\section{Chebyshev polynomials of the first kind}
\label{sec:chebyshev}

The Chebyshev polynomials $\{T_{n}\}$ are defined by the recurrence
relation
\begin{align}
\begin{aligned}
  T_{0}(x) &= 1 \\
  T_{1}(x) &= x \\
  T_{n+1}(x) &= 2xT_{n}(x) - T_{n-1}(x)
  \quad .
\end{aligned}
\label{eq:Chebyshev-recurrence}
\end{align}
The polynomials are orthogonal on the interval $[-1,1]$ with respect to
a weight
\begin{align}
  \int_{-1}^{1}\!\!\! T_{n}(x)T_{m}(x) \frac{\mathrm{d}x}{\sqrt{1 - x^{2}}}
  = \begin{cases}
    \pi           & n=m=0 \\
    \frac{\pi}{2} & n=m\neq{}0 \\
    0             & \text{else}
  \end{cases}
  \label{eq:Chebyshev-orthogonality}
\end{align}
so that we choose the basis functions $\{\phi_{\alpha}\}$ and their
duals $\{\bar{\phi}_{\alpha}\}$ on the interval $r\in{}(0, R_{\textup{c}})$
(for the radial expansion) as
\begin{align}
  \phi_{\alpha}(r)
  &= \frac{k}{2\pi\sqrt{\frac{r}{R_{\textup{c}}} - \frac{r^{2}}{R^{2}_{\textup{c}}}}}
    \,T_{\alpha}\Bigl(\frac{2r}{R_{\textup{c}}}-1\Bigr)
  \label{eq:phi}
\\
  \text{and}\quad
  \bar{\phi}_{\alpha}(r)
  &= T_{\alpha}\Bigl(\frac{2r}{R_{\textup{c}}}-1\Bigr)
  \quad\text{with}\quad 0 < r < R_{\textup{c}}
  \label{eq:phi'}
\end{align}
where $k=1/2$ for $\alpha=0$ and $k=1$ otherwise.
For the angular expansion, the appropriate interval is
$0\leq{}\theta<\pi$, so that $R_{\textup{c}}$ has to be replaced by
$\pi$ in Eqs.~\eqref{eq:phi} and~\eqref{eq:phi'}.

\end{document}